\documentclass[amsmath,amssymb,superscriptaddress,nobalancelastpage,prb,twocolumn]{revtex4}

\usepackage{graphicx}
\usepackage{epstopdf}
\usepackage{array,multirow,graphicx}
\usepackage{varioref}
\usepackage{xr-hyper}
\usepackage{xcolor}
\usepackage{nicefrac}
\usepackage{xfrac}
\definecolor{bl}{rgb}{0.0,0.2,0.6}
\usepackage{hyperref}
\hypersetup{colorlinks,linkcolor=blue,urlcolor=blue,citecolor=blue}
\usepackage{ulem}
\usepackage{braket}
\usepackage{amsmath}

 \newcommand{\bs}[1]{\ensuremath{\boldsymbol{\mathrm{#1}}}}

\newcommand{\LSCOovY}{La$_{1.78}$Sr$_{0.22}$CuO$_4$}
\newcommand{\LSCO}{La$_{2-x}$Sr$_x$CuO$_4$}
\newcommand{\Tl}{Tl$_2$Ba$_2$CuO$_{6+x}$}
\newcommand{\Bi}{(Bi,Pb)$_2$(Sr,La)$_2$CuO$_{6+x}$}
\newcommand{\Hg}{HgBa$_2$CuO$_{4+x}$}
\newcommand{\EuLSCO}{La$_{1.8-x}$Eu$_{0.2}$Sr$_x$CuO$_4$}

\newcommand{\dxy}{$d_{xy}$}
\newcommand{\dz}{$d_{z^2}$}
\newcommand{\dx}{$d_{x^2-y^2}$}

\newcommand{\dxz}{$d_{xz}$}
\newcommand{\dyz}{$d_{yz}$}
\newcommand{\ta}{$t_\alpha$}
\newcommand{\tap}{$t_\alpha'$}
\newcommand{\tapp}{$t_\alpha''$}
\newcommand{\tb}{$t_\beta$}
\newcommand{\tbp}{$t_\beta'$}
\newcommand{\tbz}{$t_{\beta z}$}
\newcommand{\tbzp}{$t_{\beta z}'$}
\newcommand{\tab}{$t_{\alpha\beta}$}

\newcommand{\ppol}{$\bar{\pi}$}
\newcommand{\spol}{$\bar{\sigma}$}
\newcommand{\EF}{$E_\mathrm{F}$}
\newcommand{\Tc}{$T_\mathrm{c}$}
\newcommand{\dA}{$d_\mathrm{A}$}

\newcommand*\xbar[1]{%
   \hbox{%
     \vbox{%
       \hrule height 0.5pt 
       \kern0.45ex
       \hbox{%
         \kern+0.1em
         \ensuremath{#1}%
         \kern+0.1em
       }%
     }%
   }%
}

\begin{document}

   
    \title{Direct Observation of Orbital Hybridisation in a Cuprate Superconductor }

   \author{C.~E.~Matt}
   
 \affiliation{Physik-Institut, Universit\"{a}t Z\"{u}rich, Winterthurerstrasse 190, CH-8057 Z\"{u}rich, Switzerland}
 \affiliation{Swiss Light Source, Paul Scherrer Institut, CH-5232 Villigen PSI, Switzerland}

       \author{D.~Sutter}
     \affiliation{Physik-Institut, Universit\"{a}t Z\"{u}rich, Winterthurerstrasse 190, CH-8057 Z\"{u}rich, Switzerland}
   \author{A.~M.~Cook}
  \affiliation{Physik-Institut, Universit\"{a}t Z\"{u}rich, Winterthurerstrasse 190, CH-8057 Z\"{u}rich, Switzerland}
  
   \author{Y.~Sassa}
  \affiliation{Department of Physics and Astronomy, Uppsala University, SE-75121 Uppsala, Sweden}
   \author{M.~M\aa nsson}
  \affiliation{KTH Royal Institute of Technology, Materials Physics, SE-164 40 Kista, Stockholm, Sweden}
   \author{O.~Tjernberg}
  \affiliation{KTH Royal Institute of Technology, Materials Physics, SE-164 40 Kista, Stockholm, Sweden}
  
   \author{L.~Das}
     \affiliation{Physik-Institut, Universit\"{a}t Z\"{u}rich, Winterthurerstrasse 190, CH-8057 Z\"{u}rich, Switzerland}
     
      \author{M.~Horio}
     \affiliation{Physik-Institut, Universit\"{a}t Z\"{u}rich, Winterthurerstrasse 190, CH-8057 Z\"{u}rich, Switzerland}
     
      \author{D.~Destraz}
     \affiliation{Physik-Institut, Universit\"{a}t Z\"{u}rich, Winterthurerstrasse 190, CH-8057 Z\"{u}rich, Switzerland}

   \author{C.~G.~Fatuzzo}
     \affiliation{Institute of Physics, \'Ecole Polytechnique Fed\'erale de Lausanne (EPFL), Lausanne CH-1015, Switzerland}

  \author{K.~Hauser}
   \affiliation{Physik-Institut, Universit\"{a}t Z\"{u}rich, Winterthurerstrasse 190, CH-8057 Z\"{u}rich, Switzerland}

 \author{M.~Shi}
 \affiliation{Swiss Light Source, Paul Scherrer Institut, CH-5232 Villigen PSI, Switzerland}

  \author{M.~Kobayashi}
 \affiliation{Swiss Light Source, Paul Scherrer Institut, CH-5232 Villigen PSI, Switzerland}

  \author{V.~N.~Strocov}
 \affiliation{Swiss Light Source, Paul Scherrer Institut, CH-5232 Villigen PSI, Switzerland}

 \author{T.~Schmitt}
 \affiliation{Swiss Light Source, Paul Scherrer Institut, CH-5232 Villigen PSI, Switzerland}

\author{P.~Dudin}
 \affiliation{Diamond Light Source, Harwell Campus, Didcot OX11 0DE, UK.}
 
 \author{M.~Hoesch}
 \affiliation{Diamond Light Source, Harwell Campus, Didcot OX11 0DE, UK.}

      \author{S.~Pyon}
\affiliation{Department of Advanced Materials, University of Tokyo, Kashiwa 277-8561, Japan}
\author{T.~Takayama}
\affiliation{Department of Advanced Materials, University of Tokyo, Kashiwa 277-8561, Japan}
\author{H.~Takagi}
\affiliation{Department of Advanced Materials, University of Tokyo, Kashiwa 277-8561, Japan}
      
    \author{O.~J.~Lipscombe}
         \affiliation{H. H. Wills Physics Laboratory, University of Bristol, Bristol BS8 1TL, United Kingdom}
         
     \author{S.~M.~Hayden}
       \affiliation{H. H. Wills Physics Laboratory, University of Bristol, Bristol BS8 1TL, United Kingdom}
     
     \author{T.~Kurosawa}
\affiliation{Department of Physics, Hokkaido University - Sapporo 060-0810, 
Japan}
 
\author{N.~Momono}
\affiliation{Department of Physics, Hokkaido University - Sapporo 060-0810, 
Japan}
\affiliation{Department of Applied Sciences, Muroran Institute of Technology, 
Muroran 050-8585, Japan}

\author{M.~Oda}
\affiliation{Department of Physics, Hokkaido University - Sapporo 060-0810, 
Japan}

  \author{T.~Neupert}
  \affiliation{Physik-Institut, Universit\"{a}t Z\"{u}rich, Winterthurerstrasse 190, CH-8057 Z\"{u}rich, Switzerland}
  
  \author{J.~Chang}
    \affiliation{Physik-Institut, Universit\"{a}t Z\"{u}rich, Winterthurerstrasse 190, CH-8057 Z\"{u}rich, Switzerland}

\maketitle

%
\textbf{ 
The minimal ingredients to explain the essential physics of  layered copper-oxide (cuprates) materials remains heavily debated.
Effective low-energy single-band models of the copper-oxygen orbitals are widely used
because there exists no strong experimental evidence supporting multi-band structures.
Here we report angle-resolved photoelectron spectroscopy experiments on 
La-based cuprates that provide direct observation of a two-band structure.
This electronic structure, qualitatively consistent with density functional theory, 
is parametrised by a two-orbital (\dx\ and \dz) tight-binding model. 
We quantify the orbital hybridisation  which  provides 
an explanation for the Fermi surface topology and the proximity of the van-Hove singularity to the Fermi level.
Our analysis leads to a unification of electronic hopping parameters for single-layer cuprates and we conclude that hybridisation, restraining $d$-wave pairing, is an important optimisation element for superconductivity. }

\begin{figure*}
 	\begin{center}
 		\includegraphics[width=1\textwidth]{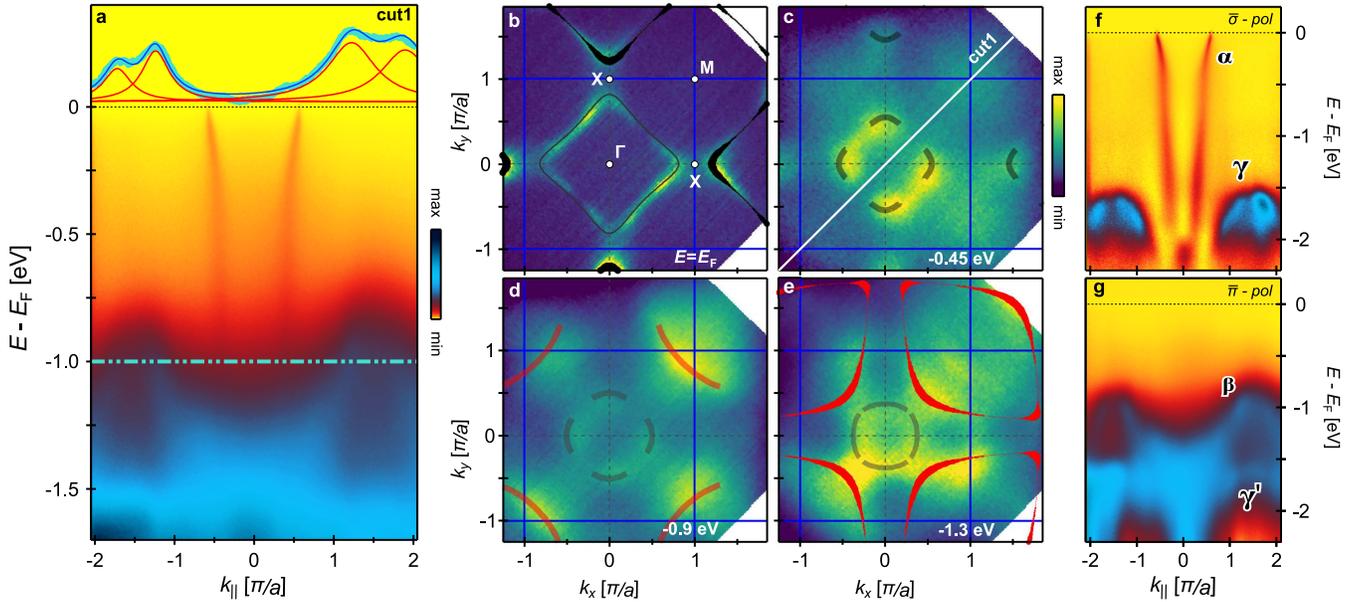}
 	\end{center}
 	\caption{\textbf{ARPES spectra showing $e_\mathrm{g}$-bands of overdoped \LSCO\ $x=0.23$.}  \textbf{(a)}, Raw ARPES energy distribution map (EDM) along cut 1 as indicated in \textbf{c}. Dashed green line indicates the position of MDC displayed on top by turquoise circles. A linear background has been subtracted from the MDC which is fitted (blue line) by four Lorentzians (red lines). 
 	\textbf{(b)}-\textbf{(e)}, Constant binding energy maps at $E_{\mathrm{F}}$ (\textbf{b}) and at higher binding energies (\textbf{c})-(\textbf{e}) as indicated. The photoemission intensity, shown in false colour scale, is integrated over $\pm10$~meV.
 	Black (red) lines indicate the position of \dx\ (\dz) bands. The curve thickness in \textbf{b} and \textbf{e} is scaled to the contribution of the \dz\ orbital. Semitransparent lines are guides to the eye. 
	\textbf{(f)}-\textbf{(g)}, EDMs along cut 1 recorded with \spol\ and \ppol\ light, (\textbf{f}) sensitive to the low-energy \dx\ and \dxz/\dyz\ bands and (\textbf{g}) the \dz\ and \dxy -derived bands. All data has been recorded with $h\nu=160$ eV.}
	 	\label{fig:fig1}
 \end{figure*}
 
Identifying the factors that limit the transition temperature \Tc\ of high-temperature cuprate superconductivity is a crucial step towards revealing the design principles underlying the pairing mechanism~\cite{LeeRMP06}. 
It may also provide an explanation for the dramatic variation of \Tc\ across the 
known single-layer compounds~\cite{PavariniPRL01}. Although superconductivity is certainly 
promoted within the copper-oxide layers,  
the apical oxygen position may play an important role in defining the transition temperature~\cite{OhtaPRB91,WeberPRL09,WeberPRB2010,HirofumiPRL10,RaimondiPRB96}. The CuO$_6$ octahedron lifts the degeneracy of the nine copper 3$d$-electrons and generates fully occupied $t_\mathrm{2g}$ and \nicefrac{3}{4}-filled $e_\mathrm{g}$ states~\cite{FinkIBM1989}. With increasing apical oxygen distance \dA\ to the CuO$_2$ plane, the $e_\mathrm{g}$ states split to create a \nicefrac{1}{2}-filled \dx\ band. 
The distance \dA\ thus defines whether single or two-band models are most appropriate to describe the low energy band structure.
It has also been predicted that \dA\ influences \Tc\ in at least two different ways.  First, the distance \dA\ controls the charge transfer gap between the oxygen and copper site which, in turn, suppresses superconductivity \cite{WeberPRB2010,  RuanScBulletin2016}.  Second, Fermi level  \dz-hybridisation, depending on \dA, reduces the pairing strength~\cite{HirofumiPRL10,HirofumiPRB12}. Experimental evidence, however, points in opposite directions. Generally, single layer materials with larger \dA\ have indeed a larger \Tc~\cite{PavariniPRL01}. However, STM studies of Bi-based cuprates suggest an anti-correlation between \dA\ and \Tc~\cite{JASlezakPNAS2008}.

In the quest to disentangle these causal relation between \dA\ and \Tc, it is imperative to experimentally 
reveal the orbital character of the cuprate band structure. 
The comparably short apical oxygen distance \dA\ makes  \LSCO\ an ideal candidate for such a study.
Experimentally, however, it is challenging to determine the orbital character of the states near the Fermi energy (\EF). In fact, the \dz\ band 
has never been identified directly by  angle-resolved photoelectron spectroscopy (ARPES) experiments. 
A large majority of ARPES studies have focused on the pseudogap, superconducting gap and quasiparticle self-energy properties in near vicinity to the Fermi level~\cite{DamascelliRMP2003}. 
An exception to this trend are studies of the so-called waterfall structure~\cite{GrafPRL2007,XiePRL2007,VallaPRL2007,MeevasanaPRB2007, ChangPRB2007} that lead to the observation of band structures below the \dx\ band~\cite{XiePRL2007,MeevasanaPRB2007}. However, the origin and hence orbital character of these bands was never addressed. Resonant inelastic x-ray scattering has been used to probe excitations between orbital $d$-levels. In this fashion, insight about the position of \dz, \dxz, \dyz, \dxy\ states with respect to \dx\ has been obtained~\cite{SalaNJP2011}. Although difficult to disentangle, it has been argued that for LSCO the \dz\ level is found above \dxz,  \dyz, \dxy~\cite{PengNatPhys2017,IvashkoPRB2017}. 
To date, a comprehensive study of the \dz\ momentum dependence is missing and therefore the coupling between the \dz\ and \dx\ bands has not been revealed. 
X-ray absorption spectroscopy (XAS) experiments, sensitive to the unoccupied states, concluded only marginal hybridisation of \dx\ and \dz\ states in \LSCO (LSCO) \cite{ChenPRL92}.
Therefore, the role of \dz-hybridisation remains ambiguous~\cite{HozoiSREP11}.      \\[2mm]

Here we provide direct ultra-violet and soft-xray ARPES measurements 
of the \dz\ band in 
La-based single layer compounds. 
The \dz\ band is located about 1~eV below the Fermi level at the Brillouin zone (BZ) corners.  From these corners, the \dz\ band is dispersing downwards along the nodal and anti-nodal directions, consistent with density functional theory (DFT) calculations.
The experimental and DFT band structure, including only \dx\ and \dz\ orbitals, is parametrised using a two-orbital tight-binding model\cite{CBishopPRB2016}.
The presence of the \dz\ band close to the Fermi level allows to describe the Fermi surface topology for all single layer compounds (including \Hg\ and \Tl) with similar hopping parameters for the \dx\ orbital. This unification of electronic parameters implies that the main difference between  single layer cuprates originates from  the hybridisation between \dx\ and \dz\ orbitals.
The significantly increased hybridisation in La-based cuprates pushes the van-Hove singularity close to the Fermi level. 
This explains why the Fermi surface differs from other single layer compounds. 
We directly quantify  the orbital hybridisation that plays a sabotaging role for superconductivity. \\[2mm] 
  
\begin{figure*}
 	\begin{center}
 		\includegraphics[width=1\textwidth]{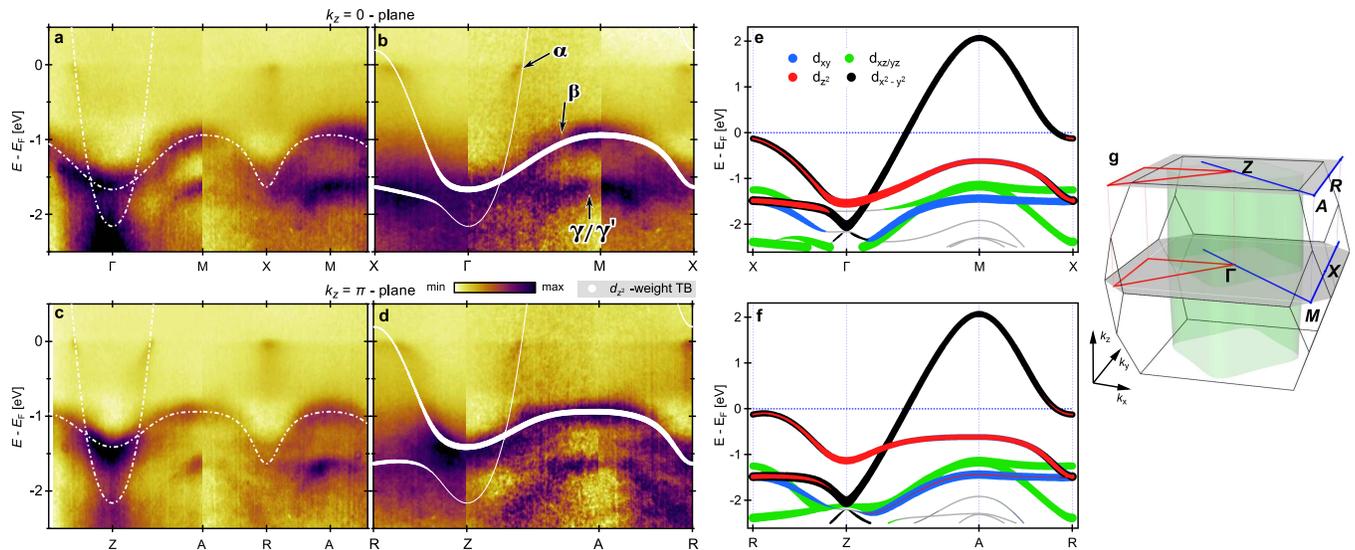}
 	\end{center}
 	\caption{\textbf{Comparison of observed and calculated band structure.} \textbf{(a)}-\textbf{(d)}, 
	Background subtracted (see methods section) soft-xray ARPES EDMs recorded on \LSCO,\ $x=0.23$
	along in-plane high-symmetry directions for $k_z=0$ and $k_z=\nicefrac{\pi}{c'}$ as indicated in \textbf{g}.
	White lines represent the two-orbital (\dz\ and \dx) tight-binding model as described in the text. The line width in \textbf{b} and \textbf{d} indicates the orbital weight of the \dz\ orbital.
	\textbf{(e)},\textbf{(f)}, Corresponding in-plane DFT band structure at $k_z=0$ and $k_z=\nicefrac{\pi}{c'}$, calculated for 
	La$_2$CuO$_4$ (see methods section).  The colour code indicates the orbital character of the bands. 
	Around the anti-nodal points ($X$ or $R$), strong hybridisation of \dz\ and \dx\ orbitals is found.
	In contrast, symmetry prevents any hybridisation along the nodal lines ($\Gamma$--$M$ or $Z$--$A$). 
	\textbf{(g)}, Sketch of the 3D BZ of LSCO with high symmetry lines and points as indicated.
	}	
		\label{fig:fig2}
 \end{figure*}

\textbf{Results}\\
\textit{Material choices:} Different dopings of \LSCO\ spanning from $x=0.12$ to 0.23 in addition 
to an overdoped compound of \EuLSCO\ with $x=0.21$ have been studied. 
These compounds represent different crystal structures:
low-temperature orthorhombic (LTO), low-temperature tetragonal (LTT)
and the high-temperature tetragonal (HTT).  Our results are very 
similar across all crystal structures and dopings (see Supplementary Fig.~1).  To keep the comparison 
to band structure calculations simple, this paper focuses on results obtained in 
the tetragonal phase of overdoped \LSCO\ with $x=0.23$.\\[2mm]

\begin{figure}
 	\begin{center}
 		\includegraphics[width=0.499\textwidth]{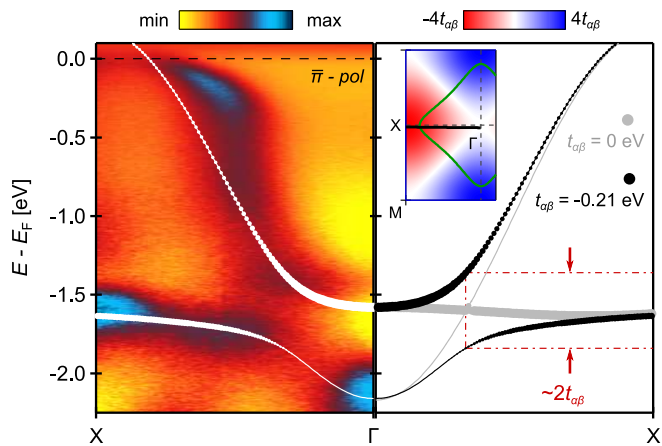}
 	\end{center}
 	\caption{\textbf{Avoided band crossing. }Left panel: Ultraviolet ARPES data recorded along the antinodal direction using 160~eV linear horizontal polarised photons. Solid white lines are the same tight-binding model as shown in Fig. 2. Right panel: Tight-binding model of the \dx\ and \dz\ bands along the anti-nodal direction. Gray lines are the model prediction in absence of inter-orbital hopping ($t_{\alpha\beta}=0$) between \dx\ and \dz. In this case, the bands are crossing near the $\Gamma$-point. This degeneracy is lifted once a finite inter-orbital hopping parameter is considered. For solid black lines $t_{\alpha\beta}=210$~meV and other hopping parameters have been adjusted accordingly. Inset indicates the Fermi surface (green line) and the $\Gamma-X$ cut directions. coloured background displays the amplitude of the hybridisation term $\Psi(\bs{k})$ that vanishes 
 on the nodal lines.  }
	 	\label{fig:fig3new}
 \end{figure}

\textit{Electronic band structure:} A raw ARPES energy distribution map (EDM), along the nodal direction, is displayed in Fig.~\ref{fig:fig1}a.
Near \EF, the widely studied nodal quasiparticle dispersion with predominately  \dx\ character is observed~\cite{DamascelliRMP2003}.  This band reveals the previously reported electron-like Fermi surface of \LSCO, $x=0.23$\cite{YoshidaPRB06,ChangNATC13} (Fig.~\ref{fig:fig1}b), the universal nodal Fermi velocity $v_{\mathrm{F}}\approx1.5$~eV\AA\cite{ZhouNAT03} and a band dispersion kink around 70 meV\cite{ZhouNAT03}. 
The main observation reported here is the second band dispersion at $\sim1$~eV below the Fermi level $E_{\mathrm{F}}$ (see Fig.~\ref{fig:fig1} and \ref{fig:fig2}) and a hybridisation gap splitting the two (see Fig.~\ref{fig:fig3new}).
 This second band -- visible in both raw momentum distribution curves (MDC) 
and constant energy maps (CEM) -- disperses downwards away from the BZ-corners. 
Since a pronounced $k_z$-dependence is observed  for this band structure (see Figs.~\ref{fig:fig2} and~\ref{fig:fig4}) a trivial surface state can be excluded . 
Subtracting a background intensity profile (see Supplementary Fig.~2) is a standard method that enhances visualisation of this second band structure. 
In fact, using soft x-rays (160 eV - 600 eV),  at least two additional bands ($\beta$ and $\gamma$) are found below the \dx\ dominated 
band crossing the Fermi level. 
Here, focus is set entirely on the $\beta$ band dispersion  closest to the \dx\ dominated band. 
This band is clearly observed at the BZ corners (see Figs.~1--3). The complete in-plane $(k_x,k_y)$ and out-of-plane ($k_z$) band dispersion is presented in Fig.~\ref{fig:fig4}. \\[2mm]


\textit{Orbital band characters:} To gain insight into the orbital character of these bands, a comparison with a DFT band structure 
calculation (see methods section) of La$_2$CuO$_4$  is shown in Fig.~\ref{fig:fig2}.
The $e_\mathrm{g}$ states (\dx\ and \dz) are generally found above the $t_\mathrm{2g}$ bands (\dxy , \dxz, and \dyz).
The overall agreement between the experiment and 
the DFT calculation (see Supplementary Fig.~3) thus suggests that the two bands nearest to the Fermi level are composed 
predominately of \dx\ and \dz\ orbitals. This conclusion can also be reached by pure 
experimental arguments.
Photoemission matrix element selection rules contain information about the orbital band character. 
They can be probed in a particular experimental setup where a mirror-plane is defined by the incident light and the electron analyser slit~\cite{DamascelliRMP2003}. 
With respect to this plane the electromagnetic light field 
has odd (even) parity for \spol\ (\ppol) polarisation (see Supplementary Fig.~4).
Orienting the mirror plane along the nodal direction (cut 1 in Fig.~\ref{fig:fig1}), the \dz\ and \dxy\ (\dx) orbitals have even (odd) parity.
For a final-state with even parity, selection rules~\cite{DamascelliRMP2003} dictate that the \dz\ and \dxy-derived bands should appear (vanish) in the \ppol\ (\spol) polarisation channel and vice versa for \dx. Due to their orientation in real-space, the  \dxz\ and  \dyz\ orbitals are not expected to show a strict switching behavior along the nodal direction~\cite{YZhangPRB2011}.
As shown in Fig.~\ref{fig:fig1}f-g, two bands ($\alpha$ and $\gamma$) appear with \spol-polarised light while for \ppol-polarised light bands $\beta$ and $\gamma'$ are observed.
Band $\alpha$ which crosses \EF\ is assigned to \dx\ while band $\gamma$ has to originate from \dxz/\dyz\ orbitals as \dz\ and \dxy-derived states are fully suppressed for \spol-polarised light. In the EDM, recorded with \ppol-polarised light, band ($\beta$) at $\sim 1$~eV binding energy and again a band ($\gamma'$) at $\sim 1.6$~eV is observed.  
From the orbital shape, a smaller $k_z$ dispersion is expected for \dx\ and \dxy-derived bands than for those from \dz\ orbitals.
As the $\beta$ band exhibits a significant $k_z$ dispersion (see Fig.~\ref{fig:fig4}), much larger than observed for 
 the \dx\ band, we conclude that it is of \dz\ character. The $\gamma'$ band which is  very close to the $\gamma$ band is  therefore of \dxy\ character.
  Interestingly, this \dz-derived band has stronger in-plane than out-of-plane dispersion, 
 suggesting that there is a significant 
  hopping to in-plane $p_x$ and $p_y$ oxygen orbitals.
Therefore the assumption that  the \dz\ states are probed uniquely through the apical oxygen $p_z$ orbital~\cite{ChenPRL92} has to be taken with caution.

\vspace{0.5 cm}
\textbf{Discussion}\\
Most minimal models aiming to describe the cuprate physics start with an approximately 
half-filled  single \dx\ band on 
a two-dimensional square lattice.
Experimentally, different band structures have been observed  across single-layer cuprate compounds.
The Fermi surface topology of 
 \LSCO\ is, for example, less rounded compared to \Bi\ (Bi2201), \Tl\ (Tl2201), and \Hg\ (Hg1201).  
 Within a single-band tight-binding model the rounded Fermi surface shape of the single layer compounds Hg1201 and Tl2201 is  described by setting $r=(|t_\alpha'|+|t_\alpha''|)/t_\alpha\sim0.4$\cite{HirofumiPRL10}, where $t_\alpha, t_\alpha'$, and $t''_{\alpha}$ are nearest, next nearest and next-next nearest neighbour hopping parameters (see Table~\ref{tab:tab1} and Supplementary Fig.~4). 
For \LSCO\ with more flat Fermi surface sections, 
significantly lower values of $r$ have been reported. For example, for overdoped \LSCOovY, 
 $r \sim0.2$ was found~\cite{YoshidaPRB06,ChangNATC13}. 
The single-band premise thus leads to varying hopping parameters across the cuprate families,
 stimulating the empirical observation that $T_\mathrm{c}^{\mathrm{max}}$ roughly scales with $t_\alpha'$~\cite{PavariniPRL01}.
This, however, is in direct contrast to $t$-$J$ models that predict the opposite correlation~\cite{WhitePRB99,MaierPRL00}.
Thus the single-band structure applied broadly to all single-layer cuprates lead to conclusions that challenge
conventional theoretical approaches. \\[2mm]

\begin{figure*}
 	\begin{center}
 		\includegraphics[width=1\textwidth]{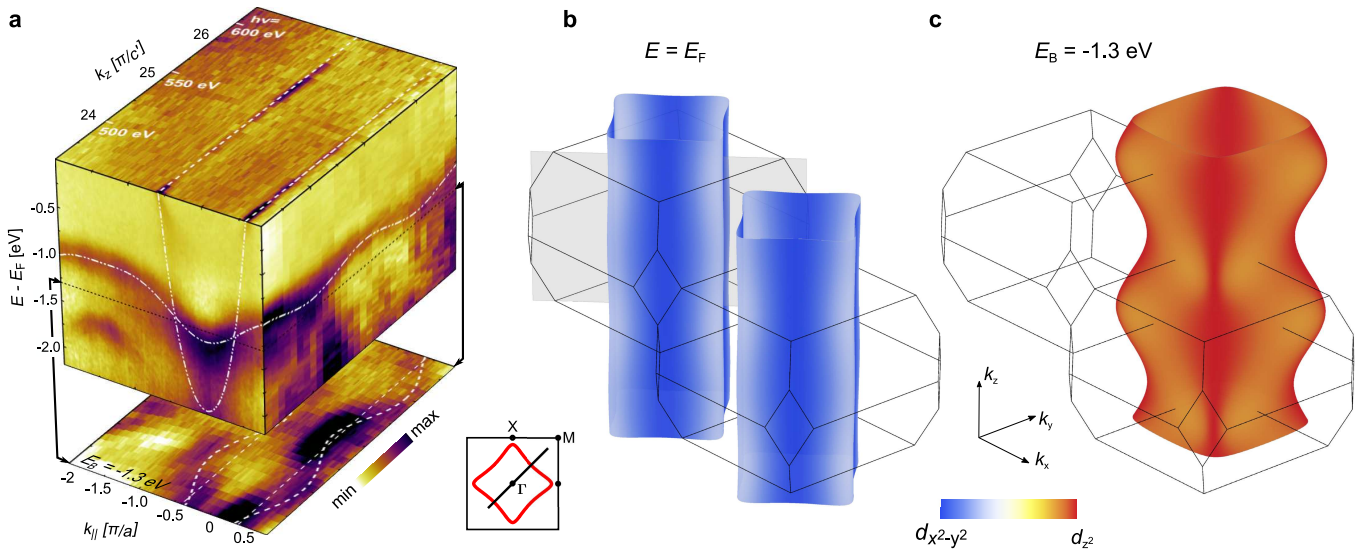}
 	\end{center}
 	\caption{\textbf{Three-dimensional band dispersion:} \textbf{(a)} $k_z$ dispersion recorded along the diagonal ($\pi,\pi$) direction of the \dx\ and \dz\ bands (along gray plane in \textbf{b}).
	Whereas the \dx\ band display no $k_z$ dependence beyond matrix element effects, the \dz\ band displays a 
	discernible $k_z$ dispersion. The iso-energy map below the cube has binding energy $E-E_{\mathrm{F}}=-1.3$~eV. White lines represent the tight-binding model.
	\textbf{(b)-(c)} Tight-binding representation of the Fermi surface ($\alpha$ band) and iso-energy  surface ($-1.3$~eV) of the $\beta$ band. The colour code indicates the $\bs{k}$-dependent orbital hybridisation. The orbital hybridisation at \EF\ is largest in the antinodal region of the $k_z=\nicefrac{\pi}{c'}$ plane where the \dz\ admixture at $k_F$  amounts to $\sim \nicefrac{1}{3}$. }	
	 	\label{fig:fig4}
 \end{figure*}
  
The observation of the \dz\ band calls for a re-evaluation of the electronic structure
in La-based cuprates using a two-orbital tight-binding model (see methods section).
Crucially, there is a hybridisation term $\Psi(\bs{k})=2 t_{\alpha\beta}[\cos( k_x )-\cos(k_y)]$ between the \dx\ and \dz\ orbitals, where $t_{\alpha\beta}$ is a hopping parameter that characterises the strength of orbital hybridisation. 
In principle, one may attempt to describe the two observed bands 
independently by taking $t_{\alpha\beta}=0$. However, the problem then
returns to the single band description with the above mentioned contradictions. 
Furthermore, $t_{\alpha\beta}=0$ implies a band
crossing in the antinodal direction that is not observed experimentally (see Fig.~\ref{fig:fig3new}). In fact, from the 
avoided band crossing one can directly estimate $t_{\alpha\beta}\approx200$~meV. 
As dictated by the different eigenvalues of the orbitals under mirror symmetry, the hybridisation term $\Psi(\bs{k})$ vanishes on the nodal lines $k_x=\pm k_y$ (see inset of Fig. 3). Hence the pure \dx\ and \dz\ orbital band character is expected along these nodal lines. 
The hybridisation $\Psi(\bs{k})$ is largest in the anti-nodal region, pushing the van-Hove singularity of the upper band close to the Fermi energy and in case of overdoped \LSCO\ across the Fermi level. 

In addition to the hybridisation parameter $t_{\alpha\beta}$ and the chemical potential $\mu$,
six free parameters 
enter the tight-binding model that yields the entire band structure (white lines in Figs.~\ref{fig:fig2} and~\ref{fig:fig4}). Nearest and next-nearest in-plane hopping
parameters between \dx\ ($t_\alpha$, $t_\alpha'$) and \dz\ ($t_{\beta},t_{\beta}'$) orbitals are introduced to capture the 
Fermi surface topology and in-plane \dz\ band dispersion (see Supplementary Fig.~4). The $k_z$ dispersion is described by nearest and next-nearest out-of-plane hoppings ($t_{\beta z}$, $t'_{\beta z}$) of the \dz\ orbital. 
The four \dz\ hopping parameters and the chemical potential $\mu$ are determined from the experimental band structure
 along the nodal direction where $\Psi(\bs{k})=0$. Furthermore, the  $\alpha$ and $\beta$ band dispersion in the anti-nodal region and the Fermi surface topology provide the parameters $t_\alpha$, $t_\alpha'$ and $t_{\alpha\beta}$. 
 Our analysis 
 reveals a finite band coupling  $-t_{\alpha\beta}=0.21$~eV 
 resulting in a strong anti-nodal orbital hybridisation (see Fig.~\ref{fig:fig2} and Table~\ref{tab:tab1}).
Compared to the single-band parametrisation \cite{YoshidaPRB06} a significantly larger value $r \sim -0.32$  is found 
and hence a unification of $t_\alpha'/t_\alpha$ ratios for all single-layer compounds is achieved.

Finally, we discuss the implication of orbital hybridisation for superconductivity and pseudogap physics.
First, we notice that a pronounced pseudogap is found in the anti-nodal region 
of \EuLSCO\ with $x=0.21$ -- consistent with transport experiments~\cite{Labiberte} (see Supplementary Fig.~5).  The fact that $t_{\alpha\beta}$ of La$_{1.59}$Eu$_{0.2}$Sr$_{0.21}$CuO$_4$ is similar to  $t_{\alpha\beta}$ of LSCO suggests that the pseudogap is not suppressed 
by the  \dz-hybridisation. 
To this end, 
a comparison to the \nicefrac{1}{4}-filled $e_\mathrm{g}$ system 
Eu$_{2-x}$Sr$_x$NiO$_4$ with $x=1.1$ is interesting~\cite{UchidaPRB11,UchidaPRL11}.
This material has the same two-orbital band structure with protection against hybridisation along 
the nodal lines. Both the \dx\ and \dz\ bands are crossing the Fermi level, producing two Fermi surface sheets~\cite{UchidaPRB11}. 
Despite an even stronger  \dz-admixture of the \dx\ derived band a  $d$-wave-like pseudogap has been reported~\cite{UchidaPRL11}. The pseudogap physics 
thus seems to be unaffected by the orbital hybridisation. 

It has been argued that orbital hybridisation -- of the kind reported here -- is unfavourable for superconducting pairing\cite{HirofumiPRL10,HirofumiPRB12}.
It thus provides an explanation for the  varying $T_{\mathrm{c}}^{\mathrm{max}}$ across single layer 
cuprate materials. Although other mechanisms, controlled by the apical oxygen distance, (e.g. variation of the copper-oxygen charge transfer gap \cite{WeberPRL09}) are not excluded our results demonstrate that orbital hybridisation exists and is an important control parameter for superconductivity.  \\[2mm]

\textbf{Acknowledgements:}
 D.S., D.D., L.D., and J.C. acknowledge support by the Swiss National Science Foundation. Further, Y.S. and M.M. are supported by the Swedish Research Council (VR) through a project (BIFROST, dnr.2016-06955).
This work was performed at the SIS\cite{SIS}, ADRESS~\cite{strocovJSYNRAD2010} and I05 beamlines at the Swiss Light Source and at the Diamond Light Source. A.M.C. wishes to thank the Aspen Center for Physics, which is supported by National Science Foundation grant PHY-1066293, for hosting during some stages of this work. 
We acknowledge Diamond Light Source for access to beamline I05 (proposal SI10550-1) that contributed to the results presented here  and thank all the beamline staff for technical support. \\

\textbf{Authors contributions:}
SP, TT, HT, TK, NM, MO, OJL,  and SMH grew and prepared single crystals. 
CEM, DS, LD, MH, DD, CGF, KH, JC, NP, MS, OT, MK, VS, TS, PD, MH, MM, and YS prepared and carried out the ARPES experiment.
CEM, KH, JC performed the data analysis. CEM carried out the DFT calculations and 
AMC, CEM, TN developed the tight-binding model.
All authors contributed to the manuscript. \\

\textbf{Methods} \\
{\it Sample characterisation:} High-quality single crystals of \LSCO,\ $x=0.12$, $0.23$, and \EuLSCO, $x=0.21$, were grown by the  floating-zone technique. 
The samples were characterised by SQUID magnetisation~\cite{LipscombePRL07} to determine superconducting transition temperatures ($T_{\mathrm{c}}=27$~K, $24$~K and $14$~K). 
For the crystal structure, the experimental lattice parameters are $a=b=3.78$~\AA\hspace{1mm}and $c=2c'=13.2$~\AA \cite{RadaelliPRB94}. \\[2mm]

{\it ARPES experiments:} Ultraviolet and soft x-ray ARPES experiments were carried out at the SIS  and ADRESS beam-lines at the Swiss Light Source and at the I05 beamline at Diamond Light Source. Samples were pre-aligned \textit{ex situ} using a x-ray LAUE instrument and cleaved  \textit{in situ} -- at base temperature (10 - 20~K) and ultra high vacuum ($\leq5\cdot10^{-11}$~mbar) -- employing a top-post technique or cleaving device\cite{cleaver}. 
Ultraviolet (soft x-ray~\cite{strocov_soft-x-ray_2014}) ARPES spectra were recorded using a SCIENTA R4000 (SPECS PHOIBOS-150) electron analyser with horizontal (vertical) slit setting. All data was recorded at the cleaving temperature 10-20~K. To visualize the \dz-dominated band, we subtracted in Figs.~\ref{fig:fig1}f,g and Figs.~\ref{fig:fig2}-\ref{fig:fig4} the background 
that was obtained by taking the minimum intensity of the MDC at each binding energy. \\[2mm]

{\it Tight-binding model:}
A two-orbital tight-binding model Hamiltonian 
with symmetry-allowed hopping terms is employed to isolate and characterise the extent of orbital hybridisation of the observed band structure \cite{CBishopPRB2016}. For compactness of the momentum-space Hamiltonian matrix representation, 
we introduce the vectors
\begin{align} \label{eq:vectors}
\bs{Q}^{\kappa}_{} &= ( a, \kappa\,  b,0)^{\mathsf{T}}, \\
\bs{R}^{\kappa_1,\kappa_2} &=  (\kappa_1 a , \kappa_1\kappa_2  b ,c )^{\mathsf{T}} /2, \nonumber\\ 
\bs{T}^{\kappa_1,\kappa_2}_1 &= (3\kappa_1 a , \kappa_1\kappa_2  b ,c )^{\mathsf{T}}/2, \nonumber\\ 
\bs{T}^{\kappa_1,\kappa_2}_2 &= (\kappa_1 a ,3 \kappa_1\kappa_2  b ,c )^{\mathsf{T}}/2, \nonumber
\end{align}

where 
$\kappa$, $\kappa_1$, and $\kappa_2$ take values $\pm1$ as defined by sums in the Hamiltonian and $\mathsf{T}$ denotes vector transposition.  

Neglecting the electron spin (spin orbit coupling is not considered) the momentum-space tight-binding Hamiltonian, $\cal{H}\left(\bs{k} \right)$, at a particular momentum $\bs{k} = \left(k_x, k_y, k_z \right)$ 
is then given by
\begin{equation} 
    \cal{H}\left(\bs{k} \right) = \left[ \begin{array}{cc}
M^{x^2-y^2}\left(\bs{k} \right)  & \Psi\left(\bs{k} \right)  \\
\Psi\left(\bs{k} \right)  & M^{z^2}\left(\bs{k} \right) 
\end{array} \right]
\label{eq: Ham}
\end{equation}
in the basis 
$\left(c_{\bs{k}, x^2 - y^2}, c_{\bs{k}, z^2} \right)^{\top}$, 
where the operator $c_{\bs{k},\alpha}$ annihilates an electron with momentum $\bs{k}$ 
in an $e_\mathrm{g}$-orbital $d_{\alpha}$, with $\alpha \in \{ x^2-y^2, z^2 \}$. 
The diagonal matrix entries
are given by
\begin{equation} \label{eq: Hxyxy}
\begin{aligned}
M^{x^2-y^2}\left(\bs{k} \right) =\hspace{1mm} & 2t_\alpha \Big[ \cos \big( k_x a\big) + \cos \big( k_y b \big) \Big] + \mu \\
&+ \sum_{\kappa=\pm1} 2 t_\alpha' \cos \left(\bs{Q}^{\kappa}_{}\cdot\bs{k} \right) \\
&+ 2t_\alpha'' \Big[ \cos \big( 2k_x a\big) + \cos \big( 2k_y b \big) \Big], 
\end{aligned}
\end{equation}
and
\begin{equation}
\begin{aligned}
M^{z^2}(\bs{k}) = \,&  2 t_{\beta} \Big[ \cos ( k_x a) + \cos ( k_y b) \Big] 
-\mu 
 \\
&+  \sum_{\kappa = \pm 1}2 t_{\beta}'\cos \left( \bs{Q}^{\kappa}_{}\cdot\bs{k} \right)  \\
&+\sum_{\kappa_{1,2} = \pm 1} \bigg[ 2 t_{\beta z} \cos \left( \bs{R}_{}^{\kappa_1,\kappa_2} \cdot\bs{k} \right)  \\
&+ \sum_{i=1,2} 2 t'_{\beta z}  \cos \left(\bs{T}^{\kappa_1,\kappa_2}_i \cdot\bs{k} \right) \bigg], \label{eq: Hzz}
\end{aligned}
\end{equation}
which describe the intra-orbital hopping for \dx\ and  $d_{z^2}$ orbitals, respectively.
The inter-orbital nearest-neighbour hopping term is given by 
\begin{equation} 
\begin{aligned}\label{eq: Hxyz}
\Psi\left(\bs{k} \right) &=\hspace{1mm} 2 t_{\alpha\beta} \Big[ \cos\big( k_x a \big) - \cos \big( k_y b \big) \Big].
\end{aligned}
\end{equation}
In the above, $\mu$ determines the chemical potential.
The hopping parameters \ta, \tap\ and \tapp\ characterise nearest neighbor (NN), next-nearest neighbour (NNN) and next-next-nearest neighbour (NNNN) intra-orbital in-plane hopping between \dx\ orbitals. \tb\ and \tbp\ characterise NN and NNN intra-orbital in-plane hopping between \dz\ orbitals, while \tbz\ and \tbzp\ characterise NN and NNN intra-orbital out-of-plane hopping between \dz\ orbitals, respectively (see Supplementary Fig.~3). Finally, the hopping parameter \tab\ characterises NN inter-orbital in-plane hopping. 
Note that in our model, \dx\ intraorbital hopping terms described by the vectors (Eqs. \ref{eq:vectors}) are neglected as these  are expected to be weak compared to those of the \dz\ orbital.
This is due to the fact that the inter-plane hopping is mostly mediated by hopping between apical oxygen $p_z$ orbitals, which in turn only hybridize with the \dz\ orbitals, not with the \dx\ orbitals. Such an argument highlights that the tight-binding model is not written in atomic orbital degrees of freedom, but in Wannier orbitals, which are formed from the Cu $d$ orbitals and the ligand oxygen $p$ orbitals. As follows from symmetry considerations and is discussed in Ref.~\onlinecite{HirofumiPRB12}, the Cu \dz\ orbital together with the apical oxygen $p_z$ orbital forms a Wannier orbital with \dz\ symmetry, while the Cu \dx\ orbital together with the four neighboring $p_\sigma$ orbitals of the in-plane oxygen forms a Wannier orbital with \dx\ symmetry. One should thus think of this tight-binding model as written in terms of these Wannier orbitals, thus implicitly containing superexchange hopping via the ligand oxygen $p$ orbitals.
Additionally we stress that all hopping parameters 
effectively include the oxygen orbitals.
Diagonalising Hamiltonian~\eqref{eq: Ham}, we find two bands 
\begin{equation}
\begin{aligned}
\varepsilon_\pm \left(\bs{k} \right) =\, & {1\over 2} \left[ M^{x^2-y^2}(\bs{k}) +  M^{z^2}(\bs{k}) \right] \\
& \pm  {1 \over 2}\sqrt{\Big[M^{x^2-y^2}(\bs{k} ) - M^{z^2}(\bs{k} )  \Big]^2 + 4 \Psi^2 (\bs{k} ) } 
\end{aligned}
\end{equation}
and make the following observations: along the  $k_x=\pm k_y$ lines, $\Psi\left(\bs{k} \right)$ vanishes and hence no orbital mixing appears in the nodal directions. 
The reason for this absence of mixing lies in the different mirror eigenvalues of the two orbitals involved. Hence it is not an artifact of the finite range of hopping processes included in our model. The parameters of the tight-binding model are determined by fitting the experimental bandstructure and are provided in Table~\ref{tab:tab1}. 
\\[2mm]

\begin{table}[ht!]
\vspace{3mm}
\begin{center}
\begin{ruledtabular}
\begin{tabular}{ccccc}
Compound & LSCO & Hg1201&Tl2201 &LSCO   \\ 
Doping $p$&0.22&0.16&0.26 &0.23\\\hline
 \multicolumn{5}{c}{Tight Binding Parameters in units of \ta$=-1.21$~eV } \\
 \hline
  $-\mu$&0.88&1.27&1.35 & 0.96\\
$-$\tap&0.13&0.47&0.42 & 0.32\\
\tapp&0.065&0.02&0.02 & 0.0\\
$-$\tab &0&0&0 & 0.175\\
\tb&-&-&- & 0.062\\
\tbp&-&-&- & 0.017\\
\tbz&-&-&- & 0.017\\
$-$\tbzp&-&-&- & 0.0017\\\hline
Ref. &\onlinecite{YoshidaPRB06}&\onlinecite{WangPRB14,VishikPRB14}& \onlinecite{PlatePRL05,PeetNJP07}&This work
\end{tabular}
\end{ruledtabular}
\caption{\textbf{Tight-binding parameters for single-layer cuprate materials.}  Comparison of tight-binding hopping parameters 
obtained from single-orbital and two-orbital models. Once a coupling $t_{\alpha\beta}$ between the \dx\ and \dz\ band is introduced for \LSCO, the \dx\ hopping parameters become comparable to those of Hg1201 and Tl2201.  }
\label{tab:tab1}	
\end{center}
\end{table}

 {\it DFT calculations:} 
Density functional theory (DFT) calculations were performed for {La$_{2}$CuO$_4$} in the tetragonal space group $I4/mmm$, No.~139, found in the overdoped regime of \LSCO\ using the WIEN2K package~\cite{Blaha2001}. 
Atomic positions are those inferred 
from neutron diffraction measurements\cite{RadaelliPRB94} for $x=0.225$. 
In the calculation, the Kohn-Sham equation is solved self-consistently by using a full-potential linear augmented plane wave (LAPW) method. The self consistent field calculation converged properly for a uniform $\bs{k}$-space grid in the irreducible BZ. The exchange-correlation term is treated within the generalized gradient approximation (GGA) in the parametrization of Perdew, Burke and Enzerhof (PBE)\cite{PerdewPRL96}. The plane wave cutoff condition was set to $RK_{\mathrm{max}}=7$ where $R$ is the radius of the smallest LAPW sphere (i.e., 1.63 times the Bohr radius) and $K_{\mathrm{max}}$ denotes the plane wave cutoff. \\[2mm]

\vspace{2mm}
\textbf{Data availability.}\\
 All experimental data are available upon request to the corresponding authors.
\vspace{2mm}

\vspace{2mm}
\textbf{Competing financial interest.}\\
The authors declare no competing financial interests.
\vspace{2mm}

\textbf{Additional information.}
Correspondence to: J.~Chang (johan.chang@physik.uzh.ch) and C.~Matt (cmatt@g.harvard.edu).

\end{document}